\documentclass[12pt,preprint]{aastex}
\usepackage{graphicx}
\usepackage[normalem]{ulem}
\newcommand\redout{\bgroup\markoverwith
{\textcolor{red}{\rule[.5ex]{2pt}{0.4pt}}}\ULon}
\usepackage{natbib}

\slugcomment{Not to appear in Nonlearned J., 45.}

\shorttitle{Partial eruption of a filament}
\shortauthors{Bi et al.}
\begin{document}
\title{
Partial eruption of a filament with twisting nonuniform fields}
\author{Yi Bi\altaffilmark{1,2},
Yunchun Jiang\altaffilmark{1},
Jiayan Yang\altaffilmark{1},
Yongyuan Xiang\altaffilmark{1},
Yunfang Cai\altaffilmark{1} AND
Weiwei Liu\altaffilmark{1}
}
\altaffiltext{1}{Yunnan Observatories, Chinese Academy of Sciences, P.O. Box 110, Kunming 650011, China; biyi@ynao.ac.cn}
\altaffiltext{2}{Key Laboratory of Solar Activity, National Astronomical Observatories, Chinese Academy of Sciences, Beijing 100012, China}
\begin{abstract}
The eruption of the filament with the kink fashion is often  regarded as a signature of the kink instability. However, the kink instability threshold for the filament magnetic structure has been not widely understood.
Using the H$\alpha$ observation from the {\it New Vacuum Solar Telescope (NVST)}, we present a partial eruptive filament. In the eruption, a filament thread appeared to split from the middle portion of the filament and to
break out in a kinklike fashion. During this period, the left filament material remained below, which erupted without the kinking motion later on.
The coronal magnetic field lines associated  with the filament are obtained from the nonlinear force-free field (NLFFF) extrapolations using the 12 minutes cadence vector
magnetograms of the Helioseismic and Magnetic Imager (HMI) on board the {\it Solar Dynamic Observatory (SDO)}.
We studied the extrapolated field lines passing through the magnetic dips that are in good agreement with the observed filament.
 The field lines are non-uniformly twisted and appear to be made up by two twisted flux ropes winding about each other. One of them has higher twist than the other, and the highly twisted one has its dips aligned with the kinking eruptive thread at the beginning of its eruption.
Before the eruption, moreover, the  highly twisted flux rope was found to expand with the approximately constant field twist.
 In addition, the helicity flux maps deduced from the HMI magnetograms show that some helicity is injected into the overlying magnetic arcade, but no significant helicity is injected into the flux ropes.
Accordingly, we suggest that the  highly twisted flux rope became kink unstable when the instability threshold declined with the expansion of the  flux rope.

\end{abstract}


\section{Introduction}

The kink instability is often considered a candidate mechanism for the eruption of the filament.
When the kink instability sets in, the flux rope's axis will experience a writhing motion due to the conservation of magnetic helicity in the highly conducting corona.
Thus, the rotation of the eruptive filament is often thought as an observational appearance of the kink instability \citep{Rust05,Green,Liu08,Bi12,Thompson,Yang12,Jiang13,Yan14}. 
Different from the rotation that is guided by surrounding magnetic field in the higher corona \citep[e.g.,][]{Cohen10,Kliem12,Bi13}, the writhing motion induced by the kink instability mainly acts in the lower corona,  below a height comparable to the footpoint distance \citep{Kliem12}.


The ideal stability of the kink mode is mainly controlled by the total twist.
The kink instability occurs if the amount of magnetic twist exceeds a critical value.
The flux tube twist is usually expressed as  ${\Phi}_{tw}=lB_{\phi}/rB_{r}$, where $B_{\phi}/B_{r}$ is the ratio of the azimuthal and axial field components of the
flux tube and $l/r$ is the length-to-width ratio of the tube.
The threshold for the onset of instability is dependant on  the detailed magnetic structure.
For a  force-free magnetic loop including the effect of photospheric line tying \citep{Hood}, the instability threshold is an amount of twist about 1.25 turns.
The MHD simulation of \citet{Fan05} showed that a line-tied flux rope becomes kink instability when the field line twist reaches about 1.7 turns.
 Consistently, the critical average twist for a magnetic loop simulated by \citet{Torok04} is about 1.75 turns.
 Moreover, the authors concluded that the instability threshold rises with the rising aspect ratio of the loop.

An inhomogeneous twisted flux rope was investigated by \citet{Birn} using magnetohydrodynamic simulations.
Their results show that a flux rope will be broken into two portions when the more strongly twisted portion becomes kink-unstable and moves rapidly outward.
The bifurcation of  a flux rope during eruption was first discussed by \citet{Gilbert} and was classed as a partial eruption by \citet{Gibson_a}, who demonstrated that part of the flux rope is expelled from the corona when it reconnects internally and with surrounding field so that it breaks in two \citep{Gibson_b}.
Their simulation implies that the writhing motion induced by the kink instability is essential for forming a current sheet within the  flux rope where it can break in two.
The observations of the partial eruption \citep[e.g.,][]{Tripathi09,Shen12,Tripathi13} exhibit that the splitting of the  flux rope is often accompanied by the writhing motion. The authors accounted for  these events using the model of \citet{Gibson_a}, but it seems that not all of these events show a clear signal of the internal magnetic reconnection.

The twisted flux rope is often found above the magnetic neutral line in the nonlinear force free magnetic field (NLFFF)  extrapolated from vector magnetograms.
Using various NLFFF extrapolation algorithms, several authors found the flux rope systems in their studied active region with twists of about 1 turn \citep{Yan01,Canou09,Inoue11,Guo13}.
Some filaments investigated \citep{Regnier,Guo10,Jing,Jiang14} were also described by the twisted flux rope with the value of the twist of approximate 1 turn, which mean that these structures are stable against kink instability. Interestingly, \citet{Guo13} reconstructed a flux rope, which has the twist of about 1.9 turn about two hours before it started to erupt. It is worth noting that the ratio of the radius  to the length of the flux rope they studied is less than 1/15, which is slightly lower than the ratio that was chosen by
\citet{Torok04} in their parameter investigation. To our knowledge, however, few investigation has focused on the evolution of the field twist associated with a filament showing the kinking motion.

 Magnetic helicity quantifies the magnetic topological complexity and is a valid tool for measuring how much a magnetic flux rope is twisted and writhed, or how much a magnetic arcade is sheared (see the review by \citet{Demoulin07} and references therein).
Inferred from the evolution of the photospheric magnetic field, magnetic helicity transported across the photosphere is often used to diagnose the helicity that is instantaneously injected into the higher solar atmosphere.
Some investigates showed that the local injection of helicity with opposite signs played a role in triggering the eruptions of the studied filaments \citep{Romano11,Dhara}.
 \citet{Romano05} found the  impulsive input of helicity at the beginning of the eruption of a kinked filament. Their results supported the idea that the amount of the magnetic helicity in the filament exceeds the limit for the kink instability  primarily due to the transport of the helicity through the photosphere.


In this article, we present the partial eruption of a filament. One portion of the filament shows the evident kinking motion in the eruption.
We present observations of the event and comparisons with the topologies of the NLFFF extrapolated field lines and the photospheric helicity injection associated with the filament. This enables us to investigate the structure and the evolution of the magnetic field relating to the kinking eruption.

\section{Observation and  data analysis}
\subsection{NVST and SDO observations}
\begin{figure*}[!ht]
\epsscale{1}
\plotone{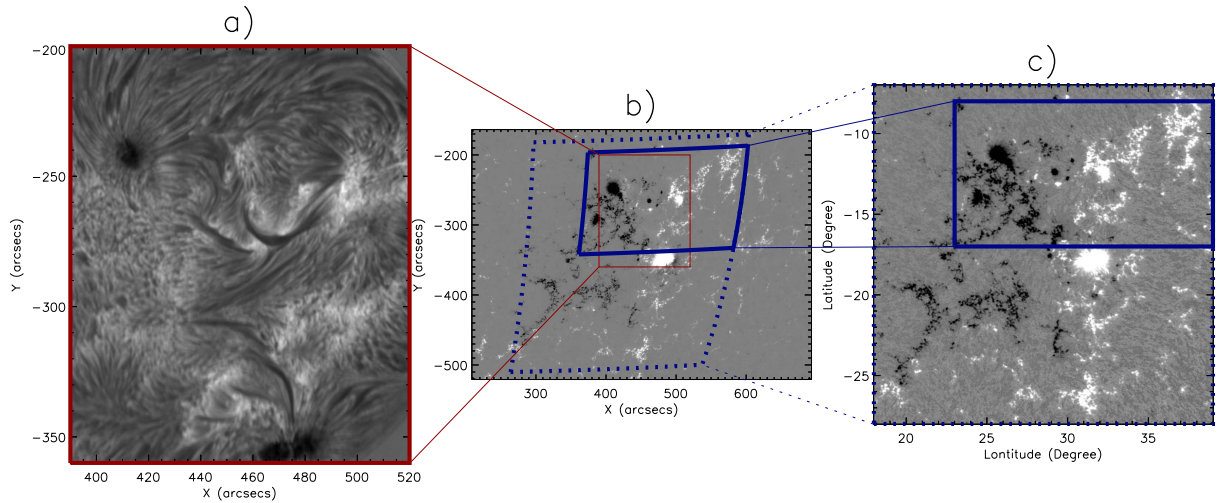}
\caption{
(a) NVST H$\alpha$ image. (b) A line-of -sight (LoS) magnetogram in the CCD coordination. (c) The radial component of the vector data in the CEA coordination.
The FOV of panel (a) and (b) is the full view of the NVST and the SHARP region, respectively.
On panel (b), the dash box colored blue indicates the position and FOV of panel (a) and the red box indicates that of panel(c)
The solid blue boxes on the panel (b) and (c) show the same position.
\label{fig1}}
\end{figure*}

Our primary data, to be used to present the eruption of the filament, are the H$\alpha$ images obtain by NVST \citep{Liu01,Liu14} and the 304 \AA\ images from the Atmospheric Imagining Assembly \citep[AIA;][]{Lemen} on board SDO.
The NVST obtains the solar images in the three channels, i.e., H$\alpha$, TiO, and G band, with a field of view (FOV) of 180$\arcsec$ $\times$ 180$\arcsec$, a cadence of 12s, and a pixel size of 0$\arcsec$.168. The observational data acquired at NVST are now available (http://fso.ynao.ac.cn).
The NVST H$\alpha$ adopted here were obtained from 04:00 UT to 07:00 UT  on 2014 Nov 4, and the whole FOV  of NVST/H$\alpha$ image  is displayed in Figure 1(a). All of the NVST images are aligned to the same FOV based on a high accuracy solar image registration procedure \citep{Feng12,Yang14}.
 The AIA takes full-disk EUV images with a pixel size of 0$\arcsec$.6 and 12s cadence.

The HMI vector field \citep{Turmon} is computed using the Very Fast Inversion of the Stokes Vector \citep[VFISV;][]{Borrero}
code and the remaining  $180^{\circ}$ azimuth ambiguity is resolved
with the Minimum Energy (ME0) code \citep{Metcalf94,Leka}.
Now, HMI provides continuous coverage of the vector field on the so-called HMI Active Region Patches (HARPs) region.
Figure 1(b) present a line-of-sight (LoS) magnetogram in the CCD coordination, with the same FOV as the HARP region  surrounding the targeted filament.
The HARP vector field data has been  remapped to a Lambert Cylindrical Equal-Area (CEA) projection and then transformed into a standard heliographic spherical coordinates.
Figure 1(c) displays the CEA map-projected field in the radial direction normal to the solar surface.


\subsection{Coronal magnetic extrapolation}
  The ``NLFFF'' package available in SSW is developed by Jim McTiernan to perform a NLFFF extrapolation using the optimization method of \citet{Wheatland00}, which is  one of the best-performing NLFF field modeling methods \citep{Schrijver08,Metcalf08}. The procedure  is fulfilled in the Cartesian and spherical coordinate and includes
the weighting function \citep{Wiegelmann04} and a preprocessing procedure to drive the observed data towards suitable boundary conditions for an extrapolation \citep{Wiegelmann06}.
In  this study, the coronal field  on the spherical geometry is extrapolated by means of the
``NLFFF'' package as follows.
 To ensure the vector data for use by NLFFF modeling efforts spans the area as large as possible \citep{Derosa}, firstly, the field covering a relatively large area  of $21^{\circ}$ $\times$ $21^{\circ}$
  (indicated by the dashed box in Figure 1(b)) is calculated within a computational domain of
 200 $\times$ 200 $\times$ 240 grids.  The CEA vector field is used as the boundary condition and the results of the potential-field source surface \citep[PFSS;][]{Schatten,Schrijver03} model as the initial condition for this process.
Secondly, the solution field is applied as the initial conditions for the further extrapolation within a domain covering an area  of $16^{\circ}$ $\times$ $9^{\circ}$ that includes the region of interest (indicated by the solid box in Figure 1(b)). The domain is resolved by  534 $\times$ 301 $\times$ 477 grids to ensure that the  size of the grid is same as the pixel size of the HMI data.
For evaluating the performance of the NLFFF extrapolation, the $\langle\theta_{i}\rangle$ metric and  $\langle|f_{i}|\rangle$ metric \citep{Wheatland00} range from  $16^{\circ}$ to $22^{\circ}$ and from $6.7\times 10^{-4}$  to $7.5\times10^{-4}$, respectively. The value is enough close to zero to ensure that the field is close to the force-free and divergence-free state.


\subsection{Twist and writhe of the magnetic field line}
 A modeled field line may indicate a thin flux tube when it is integrated with its surrounding almost parallel field. The helicity of a tube can be decomposed into contributions from twist and writhe, i.e.
 \begin{equation}\label{1}
 H=T_{w}\Phi^2+W_{r}\Phi^2
  \end{equation}
  where $\Phi$ is the magnetic flux of the tube,
the values of $T_{w}$ and $W_{r}$ are deduced from the geometry of the field lines \mbox{\citep{Berger06}}, such as
\begin{equation}\label{2}
 T_{w}=\frac{1}{2\pi}\oint{T(s)\cdot N(s)\times \frac{dN(s)}{ds}}ds
 \end{equation}
and
 \begin{equation}\label{3}
W_{r}=\frac{1}{4\pi}\oint{T(s)\times T(s')\cdot \frac{x(s)-x(s')}{|x(s)-x(s')|^{3}}}dsds'
\end{equation}
 in which the  the tangent(T) and normal(N) unit vector satisfy the Frenet-Serret formulas.

For a uniform flux rope with the twist ${\Phi}_{tw}=lB_{\phi}/rB_{r}$ (see Figure 2), $T_{w}$ is $\frac{1}{2\pi}(lB_{\phi}/r\sqrt{B_{r}^{2}+B_{\phi}^{2}})$ according to the Equation (2).
As $B_{\phi}/B_{r}$ is often smaller than 1 in the solar corona, the value of $T_{w}$ is close to that of ${\Phi}_{tw}/2\pi$ defined in the theoretical calculation. In this study, $T_{w}$ is used as the proxy for the twists of the modeled field lines.

\begin{figure*}[!ht]
\epsscale{0.5}
\plotone{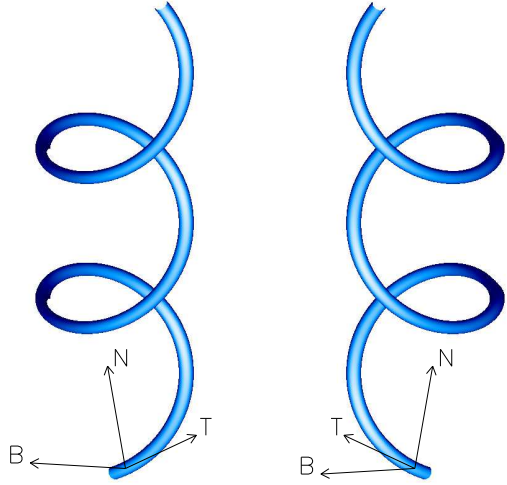}
\caption{Sketch of two field lines, which were generated using the form: $x(t)=B_{\phi}sin(t)$; $y(t)=rB_{\phi}cos(t)$; $z(t)=B_{r}t$ with $t\in[0,l/b_{r}]$, where $l/r=20$, $B_{\phi}/B_{r}=3/4$,  so that the twist ${\Phi}_{tw}=lB_{\phi}/rB_{r}=2.4\pi$, and  $lB_{\phi}/r\sqrt{B_{r}^{2}+B_{\phi}^{2}}=1.9\pi$. According to the Equation (2), $T_{w}$ for the field lines on the left (resp. right) is 1.9 (resp. -1.9). The unit vectors \^{T}, \^{N}, and \^{B}, often called the tangent, normal, and binormal unit vectors, satisfy the Frenet-Serret formulas.
\label{fig2}}
\end{figure*}



\subsection{Magnetic helicity injection}
 The rate of helicity transport across the photosphere is often computed via the proxy $G_{\theta}$ \citep{Pariat05,Dalmasse14}.
  $G_{\theta}$ is a function associated with the magnetic flux transport velocity, which is determined by local optical flow technology, such as local correlation
tracking (LCT)¡¡or differential affine velocity estimator \citep[DAVE;][]{Schuck05}.
In this study, DAVE is implemented with the window size of 20 $\times$ 20 pixels and the time interval of 720s.



It  is worth noting that the gauge invariance of the helicity flux depends on integrating over the surfaces or at least a big patch.
To better estimate the spatial distribution of the helicity flux, accordingly, \citet{Pariat05} considered  the helicity injected into an elementary flux tube, which  can be written as the  sum of the $G_{\theta}$ at the both photospheric footpoints of the corona connection.
The authors introduced a connectivity-based helicity flux density proxy, $G_{\phi}$,
such that
\begin{equation}\label{4}
  G_{\phi}(x_{\pm})=\left( G_{\theta}(x_{\pm})+G_{\theta}(x_{\mp})\left|\frac{ B_{n}(x_{\pm})}{ B_{n}(x_{\mp})}\right|  \right)f_{\pm}
\end{equation}
where $x_{+}$ and $x_{-}$ are the locations of the opposite polarities with coronal linkage, and f ranges from 0 to 1.
Then proxy $G_{\phi}$ has a meaning for the helicity flux for each individual flux tube.
Obviously, the integral of $G_{\phi}$ has the same value of that of $G_{\theta}$ over the region large enough, but $G_{\phi}$ masks some false opposite sign
signals in  $G_{\theta}$, which is canceled out in the integral of $G_{\theta}$ over the region including all magnetic polarities with coronal linkage.

\section{Results}

\begin{figure*}[!ht]
\epsscale{0.5}
\plotone{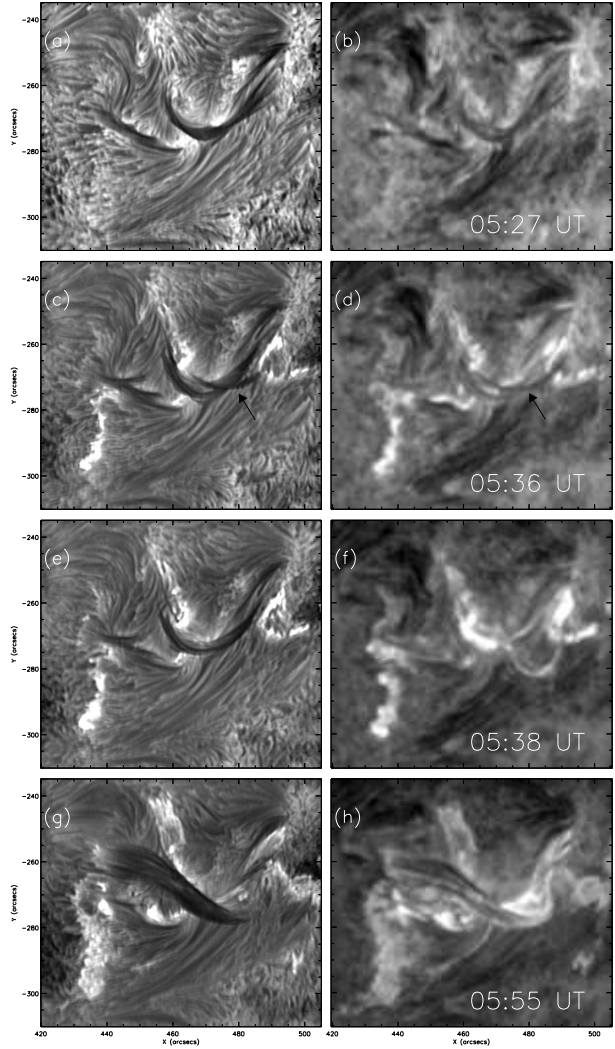}
\caption{
Left: NVST H$\alpha$ image. Right: {\it SDO} AIA 304 \AA\ images.
\label{fig3}}
\end{figure*}

 The close views of the filament on the H$\alpha$ and 304 \AA\ image are shown in Figure 3.
The filament  is located in the NOAA active region AR11884 at a position (S11, W24) on 4 October 2013. The length of the filament is approximately 30 Mm, slightly longer than that of the named mini-filament defined by \citet{Wang00}.
At 05:28 UT, a thin filament thread starts to split from the middle portion of the filament (as indicated by the arrows in Figure 3(c-d)).
The separated thread was found to rotate clockwise. Observed from the H$\alpha$ image(Figure 3(c)),  the rotation angle about  the undisturbed filament axis is approximately $30^{\circ}$ at 05:36 UT (also see the accompanying animations of Figure 3).
 At 05:38 UT, the eruptive filament thread has been disappeared from the H$\alpha$ image (Figure 3(e)) and it   evolved into a brightened one in the AIA 304 \AA\ image (Figure 3(f)).  The left portion of the filament remained undisturbed until after the previous eruptive thread has disappeared in the AIA 304 \AA\ image. Moreover, no evident rotating motion is identified in the subsequent eruption of the left portion (Figure 3(g-h)), which started about 05:45 UT.  Ultimately, a flare of X-ray class C3.2 took place with start, peak, and end times around 05:36, 05:44, and 05:52UT, respectively.


\begin{figure*}[!ht]
\epsscale{0.8}
\plotone{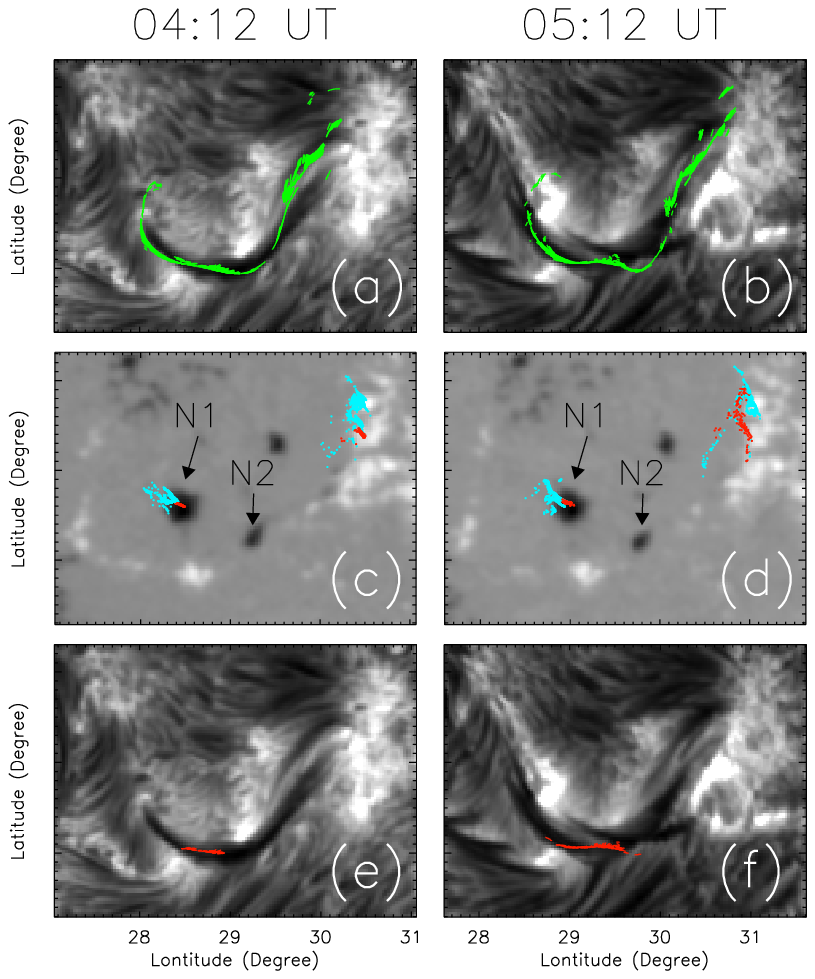}
\caption{
(a-b): The NVST/H$\alpha$ images overlaid by all of the extrapolated magnetic dips.
(c-d): The radial component of the HMI vector field overlaid by the endpoints of the fields traced from the location of the dips. The endpoints are colored red if their corresponding fields anchor the negative region having the field strength higher than 1200 G, otherwise, they are colored cyan.
(e-f): The NVST/H$\alpha$ images overlaid by the extrapolated magnetic dips, which are associated with the fields that anchor the negative region having the field strength higher than 1200 G.
 All of these images are remaped to the CEA coordination.
 (Animations of this figure are available in the online journal.)
\label{fig4}}
\end{figure*}

\begin{figure*}[!ht]
\epsscale{0.7}
\plotone{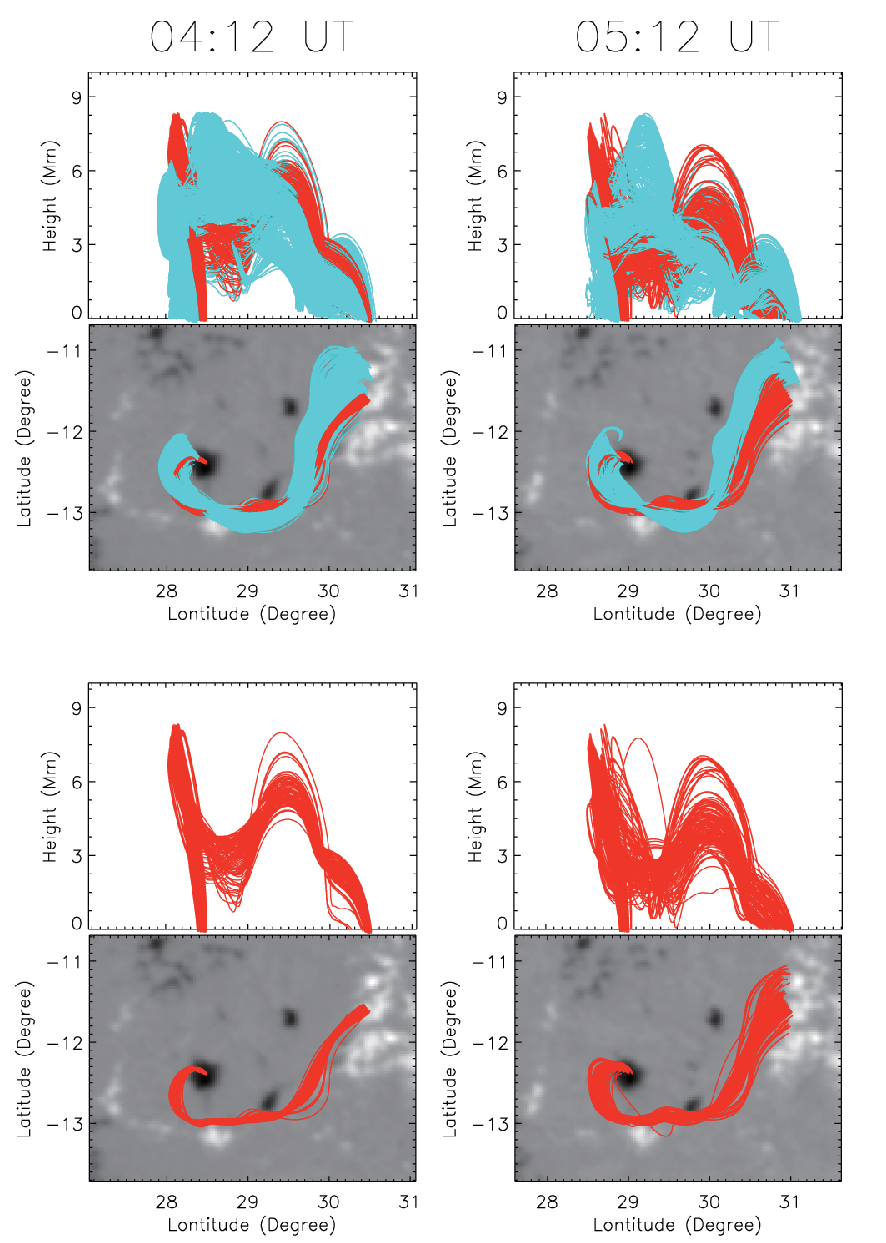}

\caption{
The field lines traced from the magnetic dips as showed in Figure 4(a-b) are displayed in the $\phi-R$ and $\phi-\theta$ plane. The fields are colored red if they anchor the negative region having the field strength higher than 1200 G, otherwise, they are colored cyan.
\label{fig5}}
\end{figure*}

 The modeled magnetic dips exist if the field lines are locally horizontal  and curved upwardly, and they are widely accepted to support the filament plasma. In Figure 4(a) and 4(b), the  green curves overlying the H$\alpha$ images show the magnetic dips calculated from the extrapolated NLFFF field at 04:12 UT and 05:12 UT, respectively. At these two moments, the locations of the dips seem to be very well aligned with the filament over its length.

The field lines traced from all of the dips are displayed in the $\phi-R$ and $\phi-\theta$ planes (Figure 5).
 After checking the character of the each field line, we assign the field lines into two groups:  the field  lines are colored red if their endpoints (as showed in Figure 4(c-d)) that anchor in negative polarity have the field strength higher than 1200 G, otherwise, the field lines are colored cyan. After such a grouping, the field lines seem to be made up by  two flux ropes winding with each other(see the First two rows of Figure 5).
Red colored field lines show a double peaks structure with their two peaks being located above the most of cyan colored field lines, while the majority of cyan field lines have their peaks lying between the two peaks of red field lines. Moreover, the dips of the red colored field lines match well the middle segment of the observed filament(Figure 4(e)). In figure 4(f), the eruptive filament thread appears to be lying across  the middle part of the filament, where approximately the dips are located.

 After obtaining the value of the twist and the writhe for each field line according to the equation (2), we found that the extrapolated field lines are non-uniformly twisted. The values of the twists range from 0.8 to 2.0 turn and average about 1.2 turn.  Furthermore, the average twist for each field line colored red is higher than the average for the field line colored cyan. The averaged value for the red (resp. cyan) is 1.55 (resp. 1.25) at 04:12 UT and 1.51 (resp. 1.1) at 05:12 UT.
The value of the writhe is smaller than that of the twist. The average writhe of the field lines colored by red has the value of  -0.04  at 04:12 UT and -0.02 at 05:12 UT.

Possessing higher twist, the flux rope colored red is extracted to be further concerned on the last two rows of Figure 5.
 The  flux rope has  the height of about 7 Mm referred to the solar surface, and has the length of about 30 Mm.
 Comparing the two panels, we can find that the flux rope evolves to be fatter. The averaged radius of the  flux rope is about 2 Mm (resp. 4 Mm) at 04:12 (resp. 05:12) UT.
 Hence, the ratio between the radius and the length of each flux ropes is about 1/15 and 1/8 at these two moments, respectively.

\begin{figure*}[ht]
\epsscale{0.65}
\plotone{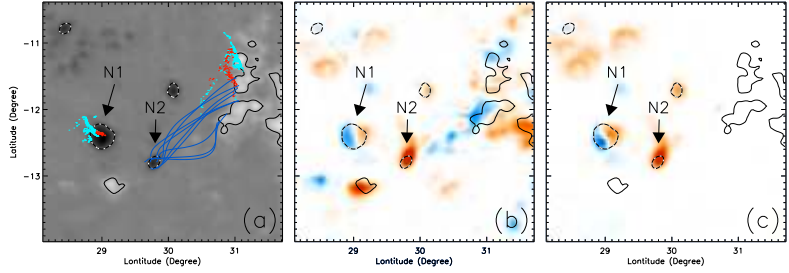}
\caption{
(a) The radial component of the HMI vector field at 05:12 UT.
(b)  The time-averaged $G_{\theta}$ map.
(c) The time-averaged $G_{\phi}$ map.
On panel (a), the selected field lines connecting the N2 are colored blue, and the endpoints of  the field lines as the Figure 4(d) showed are overlaid.
On panel (b-c), the color shows
the strength of the helicity flux with white being 0, red positive,
and blue negative.
The radial component of the HMI vector field is contoured on these panels. The contour levels are 40G for positive polarities (solid) and -40G for negative polarities (dash).
\label{fig6}}
\end{figure*}

Figure 6(b) displays the time-averaged helicity flux map computed with the proxy $G_{\theta}$ by a series of HMI magnetograms at 12 minute cadence from 04:12 UT to 05:36 UT.
 Figure 6(a) exhibits one of the magnetograms, on which the two negative polarities is referred as N1 and N2, respectively.
 From the NLFFF, we note that the N1 is where the eastern endpoint of the filament anchors and the N2 connects the field overlying the filament, while no other  negative patch flux is associated with the filament. And yet the  part of the positive polarity around the western footpoint of the filament connects the large-scale field that anchored far away from the filament.
   Accordingly, we choose f=1 to compute the $G_{\phi}(x_{-})$ (Figure 6(c)), which assumes the helicity is injected from negative polarity. In the $G_{\phi}(x_{-})$ map, therefore, helicity flux from the N1 and N2 are roughly equivalent to  the total helicity injection into the filament.

   As showed in Figure 6(b-c), both the $G_{\theta}$ and $G_{\phi}$ maps display mixed signals in N1.  However, the $G_{\phi}$ map presents the more positive flux on the right side in N1, and then the total helicity flux in N1 of  $G_{\phi}$ is close to 0.  It is consistent with the result deduced from the extrapolated field lines, which show that no increase of the magnetic helicity was discovered in the magnetic structure associated with the filament. In contrast, the
distributions of $G_{\theta}$ and $G_{\phi}$ are similar in N2, which is completely filled with positive helicity flux.

\section{Conclusions and discussion}

By means of observations from NVST and SDO, we present a partial filament eruption. The eruption starts with a thin filament thread being separated from the  filament, and the thin thread breaks out in a kinklike fashion.
We regarded the field lines with their dips aligning the observed filament as the magnetic structure associated with the filament, and investigated the twists and writhes of the field lines based on their geometries.
The result demonstrates that the magnetic system consists of two flux ropes, and one of them is estimated to be more highly twisted than the other. Moreover, the highly twisted flux rope has its dips aligned well with the thin filament thread that shows kinking motion in eruption.
  Therefore, the splitting of the eruption may result from that the more strongly twisted portion of the magnetic system becomes kink-unstable \citep{Birn}.

 For each field line constituting the highly twisted flux rope, the average twist is found to decrease slightly  with a small increase of the average writhe as time  passes. Hence, some twist may have been  quasi-statically converted into the writhe during this period. A similar equilibrium configuration was obtained in the MHD simulations of \citet{Fan05}, who showed a mildly kinked equilibrium with some finite writhing of the flux rope axis. Moreover, the helicity for the highly twisted flux rope, which is equivalent to the sum of the values of the twist and the writhe according to the Equation (1), amounts to $1.51\Phi^{2}$ and   $1.49\Phi^{2}$ at 04:12 UT and 05:12 UT, respectively. It indicates that no increase of the amount of the magnetic helicity was detected in the flux rope before its eruption. In this case, the question is the way in which the flux rope develops to be nonlinear kink instability as implied by the observed strong rotation of the filament thread.

As mentioned in the Section 2, the value of $T_{w}$ is slightly smaller than that of the twist ${\Phi}_{tw}=lB_{\phi}/rB_{r}$ for the same flux rope. It implies that the twist ${\Phi}_{tw}$ for our modeled flux rope should be slightly greater than 1.5 turns that is deduced from the proxy $T_{w}$.
  \citet{Torok04} estimated ${\Phi}_{tw}=1.75$ turn as the threshold of the kink-instability of a magnetic loop with their defined loop aspect ratio of about 5,
which amounts to a ratio of 11.2 between the width and the length of the loop.
In our extrapolation, the width-length ratio of the modeled flux rope increase from 1/15 to 1/8 before the eruption. i.e. from less than to greater than that given by \citet{Torok04}.
 As the instability threshold rises with rising aspect ratio \citep{Torok04}, it is possible that the threshold for the kink instability is greater than 1.75 turn at 04:12 UT and is less than 1.75 turn at 05:12 UT in our extrapolation.
Hence, the field structure may become kink unstable when it expands with the approximate  constant of the twist.

   On the other hand, the helicity flux from the photosphere is obtained using both the proxies $G_{\theta}$ and $G_{\phi}$. Since the proxy $G_{\phi}$  involves field connectivity of each elementary flux tube,  the proxy $G_{\phi}$ is better than the proxy  $G_{\theta}$ to present the distribution of photospheric helicity flux density per flux tube.  However, direct interpretation of the distribution of the helicity flux should be made with caution, because the determination of the helicity flux requires  the gauge selection for the definition of the vector potential $A(\nabla \times A =B)$ and both
 the helicity proxies $G_{\theta}$ and $G_{\phi}$ are calculated with the classical gauge of $\nabla \times A_{p} = 0$ and $A_{p} \cdot n = 0 $.
 By contrast, there is no gauge involved in $T_{w}$ and $W_{r}$ according to the Equation (2) and (3). In our study, the results from both methods show that  no significant magnetic helicity is injected into the magnetic  flux rope passing through the magnetic dips.

   Moreover, both proxies $G_{\theta}$ and $G_{\phi}$ exhibit that some positive helicity is injected from  a patch flux (referred as N2) which connects the magnetic arcade overlying the studied filament. The local injection of the helicity may play a role in twisting the overlying arcades.   If the overlying field is weakened by the twisting, as suggested by \citet{Torok13}, the underlying  flux rope would undergo a quasi-static expansion. Therefore, the helicity injection into the overlying arcade may have a role in expanding the  flux rope as showed in the extrapolation.




\acknowledgments
The authors sincerely thank the anonymous referee for his/her detailed comments and useful suggestions to improve this manuscript.
This work is supported by the Natural Science Foundation of China under grants 11403098, 11173058, 11473065, and 11273056, and by the Open
Research Programs of CAS.
The NASA/SDO data used here are courtesy of the
  AIA and HMI science teams.  The H$\alpha$  data used in this paper were obtained with the New Vacuum Solar Telescope in Fuxian Solar Observatory of Yunnan Astronomical Observatory, CAS.

\nocite{*}


\begin{thebibliography}{plainnat}
\bibitem[Berger \& Prior(2006)]{Berger06} Berger, M.~A., \& Prior, C.\ 2006, Journal of Physics A Mathematical General, 39, 8321
\bibitem[Bi et al.(2012)]{Bi12} Bi, Y., Jiang, Y., Li, H.,
Hong, J., \& Zheng, R.\ 2012, \apj, 758, 42
\bibitem[Bi et al.(2013)]{Bi13} Bi, Y., Jiang, Y., Yang, J.,
et al.\ 2013, \apj, 773, 162


\bibitem[Birn et al.(2006)]{Birn} Birn, J., Forbes, T.~G., \& Hesse, M.\ 2006, \apj, 645, 732
\bibitem[Borrero et al.(2011)]{Borrero} Borrero, J.~M.,
Tomczyk, S., Kubo, M., et al.\ 2011, \solphys, 273, 267


\bibitem[Canou et al.(2009)]{Canou09} Canou, A., Amari, T., Bommier, V., et al.\ 2009, \apjl, 693, L27
\bibitem[Cohen et al.(2010)]{Cohen10} Cohen, O., Attrill, G.~D.~R., Schwadron, N.~A., et al.\ 2010, Journal of Geophysical Research (Space Physics), 115, A10104
\bibitem[Dalmasse et al.(2014)]{Dalmasse14} Dalmasse, K., Pariat,
E., D{\'e}moulin, P., \& Aulanier, G.\ 2014, \solphys, 289, 107
\bibitem[D{\'e}moulin(2007)]{Demoulin07} D{\'e}moulin, P.\ 2007, Advances in Space Research, 39, 1674

\bibitem[De Rosa et al.(2009)]{Derosa} De Rosa, M.~L., Schrijver, C.~J., Barnes, G., et al.\ 2009, \apj, 696, 1780
\bibitem[Dhara et al.(2014)]{Dhara} Dhara, S.~K., Ravindra, B., \& Banyal, R.~K.\ 2014, \na, 26, 86
\bibitem[Fan(2005)]{Fan05} Fan, Y.\ 2005, \apj, 630, 543
\bibitem[Feng et al.(2012)]{Feng12} Feng, S, Deng, L.~H., Shu, G.~F., et al.\ 2012, 2012 IEEE Fifth International Conference on Advanced Computational Intelligence (ICACI), October 18-20, 2012 Nanjing, Jiangsu, China
\bibitem[Gibson \& Fan(2006)]{Gibson_a} Gibson, S.~E., \& Fan, Y.\ 2006, \apjl, 637, L65
\bibitem[Gibson \& Fan(2006)]{Gibson_b} Gibson, S.~E., \& Fan, Y.\ 2006, Journal of Geophysical Research (Space Physics), 111, A12103
\bibitem[Gilbert et al.(2000)]{Gilbert} Gilbert, H.~R., Holzer, T.~E., Burkepile, J.~T., \& Hundhausen, A.~J.\ 2000, \apj, 537, 503
\bibitem[Green et al.(2007)]{Green} Green, L.~M., Kliem, B., T{\"o}r{\"o}k, T., van Driel-Gesztelyi, L., \& Attrill, G.~D.~R.\ 2007, \solphys, 246, 365
   \bibitem[Guo et al.(2010)]{Guo10} Guo, Y., Schmieder, B., D{\'e}moulin, P., et al.\ 2010, \apj, 714, 343
\bibitem[Guo et al.(2013)]{Guo13} Guo, Y., Ding, M.~D., Cheng, X., Zhao, J., \& Pariat, E.\ 2013, \apj, 779, 157

\bibitem[Hood \& Priest(1981)]{Hood} Hood, A.~W., \& Priest, E.~R.\ 1981, Geophysical and Astrophysical Fluid Dynamics, 17, 297
\bibitem[Inoue et al.(2011)]{Inoue11} Inoue, S., Kusano, K., Magara, T., Shiota, D., \& Yamamoto, T.~T.\ 2011, \apj, 738, 161
\bibitem[Isenberg \& Forbes(2007)]{Isenberg07} Isenberg, P.~A., \& Forbes, T.~G.\ 2007, \apj, 670, 1453

\bibitem[Jiang et al.(2014)]{Jiang14} Jiang, C., Wu, S.~T., Feng, X., \& Hu, Q.\ 2014, \apjl, 786, LL16
\bibitem[Jiang et al.(2013)]{Jiang13} Jiang, Y., Hong, J., Yang, J., et al.\ 2013, \apj, 764, 68

\bibitem[Jing et al.(2014)]{Jing} Jing, J., Liu, C., Lee,
J., et al.\ 2014, \apjl, 784, LL13
\bibitem[Kliem et al.(2012)]{Kliem12} Kliem, B., T{\"o}r{\"o}k,
T., \& Thompson, W.~T.\ 2012, \solphys, 281, 137

\bibitem[Leka et al.(2009)]{Leka} Leka, K.~D., Barnes, G.,
Crouch, A.~D., et al.\ 2009, \solphys, 260, 83
\bibitem[Lemen et al.(2012)]{Lemen} Lemen, J.~R., Title,
A.~M., Akin, D.~J., et al.\ 2012, \solphys, 275, 17
\bibitem[Liu
\& Beckers(2001)]{Liu01} Liu, Z., \& Beckers, J.~M.\ 2001, \solphys, 198, 197
\bibitem[Liu(2008)]{Liu08} Liu, Y.\ 2008, \apjl, 679, L151
\bibitem[Liu et al.(2014)]{Liu14} Liu, Z., Xu, J., Gu, B.-Z.,
et al.\ 2014, Research in Astronomy and Astrophysics, 14, 705



\bibitem[Metcalf(1994)]{Metcalf94} Metcalf, T.~R.\ 1994,
\solphys, 155, 235
\bibitem[Metcalf et al.(2008)]{Metcalf08} Metcalf, T.~R., De Rosa, M.~L., Schrijver, C.~J., et al.\ 2008, \solphys, 247, 269
\bibitem[Pariat et
al.(2005)]{Pariat05} Pariat, E., D{\'e}moulin, P., \& Berger, M.~A.\ 2005, \aap, 439, 1191

\bibitem[R{\'e}gnier \& Amari(2004)]{Regnier} R{\'e}gnier, S., \& Amari, T.\ 2004, \aap, 425, 345
\bibitem[Romano et al.(2011)]{Romano11} Romano, P., Pariat, E., Sicari, M., \& Zuccarello, F.\ 2011, \aap, 525, AA13
\bibitem[Romano et al.(2005)]{Romano05} Romano, P., Contarino, L., \& Zuccarello, F.\ 2005, \aap, 433, 683
\bibitem[Rust \& LaBonte(2005)]{Rust05} Rust, D.~M., \& LaBonte, B.~J.\ 2005, \apjl, 622, L69
\bibitem[Schatten et al.(1969)]{Schatten} Schatten, K.~H.,
Wilcox, J.~M., \& Ness, N.~F.\ 1969, \solphys, 6, 442
\bibitem[Schrijver
\& De Rosa(2003)]{Schrijver03} Schrijver, C.~J., \& De Rosa, M.~L.\ 2003, \solphys, 212, 165
\bibitem[Schrijver et al.(2006)]{Schrijver08} Schrijver, C.~J., De Rosa, M.~L., Metcalf, T.~R., et al.\ 2006, \solphys, 235, 161
\bibitem[Shen et al.(2012)]{Shen12} Shen, Y., Liu, Y.,
\& Su, J.\ 2012, \apj, 750, 12
\bibitem[Schuck(2005)]{Schuck05} Schuck, P.~W.\ 2005, \apjl,
632, L53



\bibitem[Thompson et al.(2012)]{Thompson} Thompson, W.~T., Kliem, B., T{\"o}r{\"o}k, T.\ 2012, \solphys, 276, 241
\bibitem[T{\"o}r{\"o}k et al.(2004)]{Torok04} T{\"o}r{\"o}k, T., Kliem, B., \& Titov, V.~S.\ 2004, \aap, 413, L27
\bibitem[T{\"o}r{\"o}k et al.(2013)]{Torok13} T{\"o}r{\"o}k, T., Temmer, M., Valori, G., et al.\ 2013, \solphys, 286, 453
\bibitem[Tripathi et al.(2009)]{Tripathi09} Tripathi, D., Gibson, S.~E., Qiu, J., et al.\ 2009, \aap, 498, 295
\bibitem[Tripathi et al.(2013)]{Tripathi13} Tripathi, D., Reeves, K.~K., Gibson, S.~E., Srivastava, A., \& Joshi, N.~C.\ 2013, \apj, 778, 142
\bibitem[Turmon et al.(2010)]{Turmon} Turmon, M., Jones, H.~P., Malanushenko, O.~V., \& Pap, J.~M.\ 2010, \solphys, 262, 277


\bibitem[Wang et al.(2000)]{Wang00} Wang, J., Li, W., Denker,
C., et al.\ 2000, \apj, 530, 1071
\bibitem[Wheatland et al.(2000)]{Wheatland00} Wheatland, M.~S., Sturrock, P.~A., \& Roumeliotis, G.\ 2000, \apj, 540, 1150
\bibitem[Wiegelmann(2004)]{Wiegelmann04} Wiegelmann, T.\ 2004, \solphys, 219, 87

\bibitem[Wiegelmann et al.(2006)]{Wiegelmann06} Wiegelmann, T., Inhester, B., \& Sakurai, T.\ 2006, \solphys, 233, 215
\bibitem[Yan et al.(2001)]{Yan01} Yan, Y., Deng, Y., Karlick{\'y}, M., et al.\ 2001, \apjl, 551, L115
\bibitem[Yang et al.(2012)]{Yang12} Yang, J., Jiang, Y., Bi,
Y., et al.\ 2012, \apj, 749, 12
\bibitem[Yang et al.(2014)]{Yang14} Yang, Y.-F., Qu, H.-X., Ji, K.-F., et al.\ 2014, arXiv:1407.7958
\bibitem[Yan et al.(2014)]{Yan14} Yan, X.~L., Xue, Z.~K.,
Liu, J.~H., et al.\ 2014, \apj, 782, 67

\end{thebibliography}
\end{document}